\documentclass[trackchanges,twocolumn]{aastex7}
\usepackage{amsmath}

\renewcommand{\today}{}  
\preprint{}

\begin{document}

\title{Imprints of Different Types of Low-Angular-Momentum Accretion Flow Solutions in General Relativistic Hydrodynamic Simulations}

\author[orcid=0000-0002-4064-0446,gname=Indu K. ,sname='Dihingia']{Indu  K. Dihingia}
\affiliation{Tsung-Dao Lee Institute, Shanghai Jiao Tong University, 1 Lisuo Road, Shanghai, 201210, People’s Republic of China}
\email[show]{ikd4638@gmail.com, ikd4638@sjtu.edu.cn}

\author[orcid=0000-0001-8213-646X,gname=Uniyal, sname='Akhil']{Akhil Uniyal} 
\affiliation{Tsung-Dao Lee Institute, Shanghai Jiao Tong University, 1 Lisuo Road, Shanghai, 201210, People’s Republic of China}
\email{akhil_uniyal@sjtu.edu.cn}

\author[orcid=0000-0002-8131-6730,gname=Yosuke, sname='Mizuno']{Yosuke Mizuno}
\affiliation{Tsung-Dao Lee Institute, Shanghai Jiao Tong University, 1 Lisuo Road, Shanghai, 201210, People’s Republic of China}
\affiliation{School of Physics and Astronomy, Shanghai Jiao Tong University, 800 Dongchuan Road, Shanghai, 200240, People’s Republic of China}
\affiliation{Key Laboratory for Particle Physics, Astrophysics and Cosmology (MOE), Shanghai Key Laboratory for Particle Physics and Cosmology, Shanghai Jiao-Tong University,800 Dongchuan Road, Shanghai, 200240, People's Republic of China}
\affiliation{Institut f\"{u}r Theoretische Physik, Goethe-Universit\"{a}t Frankfurt, Max-von-Laue-Strasse 1, D-60438 Frankfurt am Main, Germany}
\email{mizuno@sjtu.edu.cn}
 
\begin{abstract}
Depending on the astrophysical source and its environment, the accretion flows can exhibit a variety of behaviors and characteristics in accordance with the type of solutions. We study low-angular-momentum accretion flows onto black holes using two-dimensional general relativistic hydrodynamic (GRHD) simulations to find imprints of different types of accretion solutions. Such flows, relevant to X-ray binaries and wind-fed low-luminosity active galactic nuclei, often lack sufficient angular momentum to form standard accretion disks. We initialize simulations with semi-analytical transonic solutions defined by specific energy (${\cal E}_0$) and angular momentum ($\lambda_0$), allowing a systematic classification of flow types with: (i) an outer sonic point, (ii) an inner sonic point, and (iii) both, exhibiting shock transitions. Only solutions with two sonic points produce hot, thermally driven bipolar jets/outflows with Lorentz factors up to $\gamma\sim2$, despite the absence of magnetic fields. Using a general relativistic radiation transfer calculation, we compute broadband spectra and images at X-ray ($1 \, \rm keV$) from bremsstrahlung emission. Radiative properties depend strongly on the type of accretion solution. Solutions with inner sonic points produce the brightest and most extended X-ray emission, while outer-point solutions produce compact, fainter signals. These multidimensional models are thus essential for predicting radiative signatures and will enable the development of semi-analytical tools for interpreting X-ray binaries and possibly Sgr~A$^*$ in weak magnetic field regimes.
\end{abstract}


\section{Introduction} 
Some of the most dynamic and luminous events in the universe are powered by accretion onto compact objects such as black holes and neutron stars~\citep{Frank-etal2002}. Although the transport of mass and angular momentum within accretion disks has been the focus of extensive theoretical and numerical work (see~\citealt{Davis-Tchekhovskoy2020, Mizuno2022, Dihingia-Fendt2024} for reviews), comparatively less attention has been given to the processes governing the formation of these disks. A major difficulty arises from the large-scale gap between the size of the compact object and the circularization radius of the infalling material, which poses significant challenges for resolving the full dynamical range in simulations. However, in astrophysical systems where the accreted gas has relatively low angular momentum, this difficulty is less severe, enabling more tractable modeling of disk formation. Such conditions are particularly relevant in systems like black hole X-ray binaries (BH-XRBs)~\citep{Sukova:2014fma, Meyer-Hofmeister:2014hsa} and in the chaotic, wind-fed accretion environments around supermassive black holes in low-luminosity active galactic nuclei, including Sgr~A$^*$ at our Galactic center~\citep{Murchikova-etal2022, Ressler-etal2023}.

Recent studies increasingly support the view that the Sgr~A$^*$ is primarily fed by stellar winds from nearby massive stars orbiting at parsec-scale distances~\citep{Quataert2004, Cuadra-etal2008, Ressler-etal2018}. The formation of an accretion disk in this context is not guaranteed and depends sensitively on both the initial conditions of the stellar winds and their collective hydrodynamic interactions~\citep{Moscibrodzka:2006pm, Shcherbakov:2010yq}. Unlike the well-ordered, rotation-supported tori typically produced in idealized accretion models, stellar-wind-fed flows in low-luminosity active galactic nuclei (AGNs) are highly irregular. These flows are often characterized by a clumpy, turbulent structure and a broad angular momentum distribution. Due to the initial low angular momentum of the matter, it has a relatively short infall timescale, and therefore, the gas generally fails to circularize before being accreted and remains largely unbound~\citep{Ressler-etal2018}. Accretion in such environments is dominated by the direct infall of low-angular-momentum material rather than by the gradual viscous evolution of a coherent disk.

Simulating accretion flows from large-scale feeding down to the event horizon presents significant computational challenges, primarily due to the vast range of spatial and temporal scales involved. As a result, an alternative and complementary approach is to investigate transonic accretion solutions that extend from the black hole horizon to large distances. This strategy mirrors the way analytic models of circular fluid motion have historically advanced our understanding of accretion disk structure. In the context of low-angular-momentum accretion, analytic studies have shown that the nature of the flow depends sensitively on the energy and angular momentum of the infalling gas~\citep{Fukue1987, Chakrabarti1989, Chakrabarti1996, Chakrabarti:2004uy}. These solutions reveal a variety of possible flow regimes, including smooth, quasi-spherical Bondi-like inflow~\citep{Bondi1952}, flows containing shocks~\citep[e.g.,][]{Fukue1987, Chakrabarti1989}. 
There have been substantial efforts to study low-angular momentum flow in semi-analytical or simulation fronts, finding their applications in different astrophysical scenarios \citep[e.g.,][]{Molteni-etal1994,Ryu-etal1995,Molteni-etal1996a,Molteni-etal1996b,Lanzafame-etal1998,Proga-Begelman2003, Chakrabarti-etal2004,Giri-etal2010,Okuda-Molteni2012, Okuda2014,Okuda-Das2015,Kim-etal2017,Okuda-etal2019,Kim-etal2019,Sukova-etal2017,Palit-etal2019,Singh-etal2021,Okuda-etal2022,Garain-Kim2023,Huang-Singh2025}. Recently some general relativistic hydrodynamic and magnetohydrodynamic simulations also have been performed~\citep{Olivares-etal2023,Dihingia-Mizuno2024,Dihingia-Mizuno2025}. The numerical simulations among these either deal with solutions only with the inner sonic points or supersonic injection of matter. Thus, the earlier simulations lack rigor in getting different types of low-angular momentum accretion solutions consistently as the analytical solutions.  

Therefore, in this study, we initialize our general relativistic hydrodynamic (GRHD) simulations using analytically derived one-dimensional transonic solutions as initial conditions \cite{Dihingia-etal2018a, Dihingia-etal2019a}. This approach allows us to self-consistently bridge large-scale inflow properties with horizon-scale dynamics, providing a physically motivated framework for exploring the evolution of low-angular-momentum accretion flows in a relativistic regime. The goal of this work is primarily to find different types of low-angular-momentum accretion solutions and search for radiative imprints of them. 

In the next section, we describe the numerical setup, and in subsequent sections, we discuss our results investigating flow properties and radiative properties. Finally, in section~8, we display our conclusions and add discussions based on the simulation results. 

\section{Numerical setup}
We investigate low-angular-momentum accretion flows using 2D ideal GRHD simulations with the \texttt{BHAC} code \citep{Porth-etal2017, Olivares-etal2019} in modified Kerr-Schild coordinates. Simulations use spherical polar coordinates $(r, \theta)$ with logarithmic radial spacing up to $2500\,r_g$, in units where $G = M_{\rm BH} = c = 1$. With this, all length scales and time scales are presented in terms of $r_g=GM_{\rm BH}/c^2$ and $t_g=GM_{\rm BH}/c^3$, respectively. The simulation domain is resolved with an effective resolution of $512\times 256$ with two levels of static mesh refinement (SMR) (base resolution $256\times 128$), where maximum SMR levels are employed around the equatorial plane ($\pm 45^\circ$). Additionally, to validate our 2D simulations, we also perform a three-dimensional (3D) simulation, considering numerical resolution $256\times128\times64$ as a test of the 3D effect.

The initial conditions, viz., four-velocities ($u^\mu$) and density ($\rho$), and pressure $(p)$ are calculated considering $\lambda_0=-u_\phi/u_t={\rm constant}$, and ${\cal E}_0=-hu_t={\rm constant}$, respectively, where $\lambda_0$ and ${\cal E}_0$ are known as the specific angular momentum and specific energy of the fluid element. The radial four-velocity ($u^r$) is obtained from semi-analytical solutions for the given $\lambda_0$ and ${\cal E}_0$. With that, all other initial quantities can be calculated. The explicit expressions of them can be seen in Appendix~\ref{App-A}. The initial gas pressure is calculated considering adiabatic approximation with an index $\Gamma=4/3$, i.e. $p= \kappa \rho^\Gamma$, where $\kappa$ is a constant related to entropy. 
During time evolution, we use an ideal equation of state, where enthalpy is given by $h=1 + \Gamma/(\Gamma-1)~p/\rho$. In the calculation, $g_{\mu\nu}$ corresponds to the metric components of the Kerr black hole in Boyer-Lindquist coordinates. For this study, we consider a black hole with spin parameter $a_*=0.9375$. 

Note that we evolve the simulations for a very long time, $t=50\,000\,t_g$, so that the impact of initial density distribution can be minimized. During this time, our simulations reach quasi-steady states. The time-averaged results are shown within $t=40\,000$-$50\,000\,t_g$, which is applied to all the analysis (Fig.~\ref{fig:01}, \ref{fig:02A}, \ref{fig:02}, \ref{fig:03} and Fig.~\ref{fig:04}, except Fig.~\ref{fig:05}).

\section{Types of solutions}
\begin{figure*}[ht]
    \centering
    \includegraphics[width=1.0\linewidth]{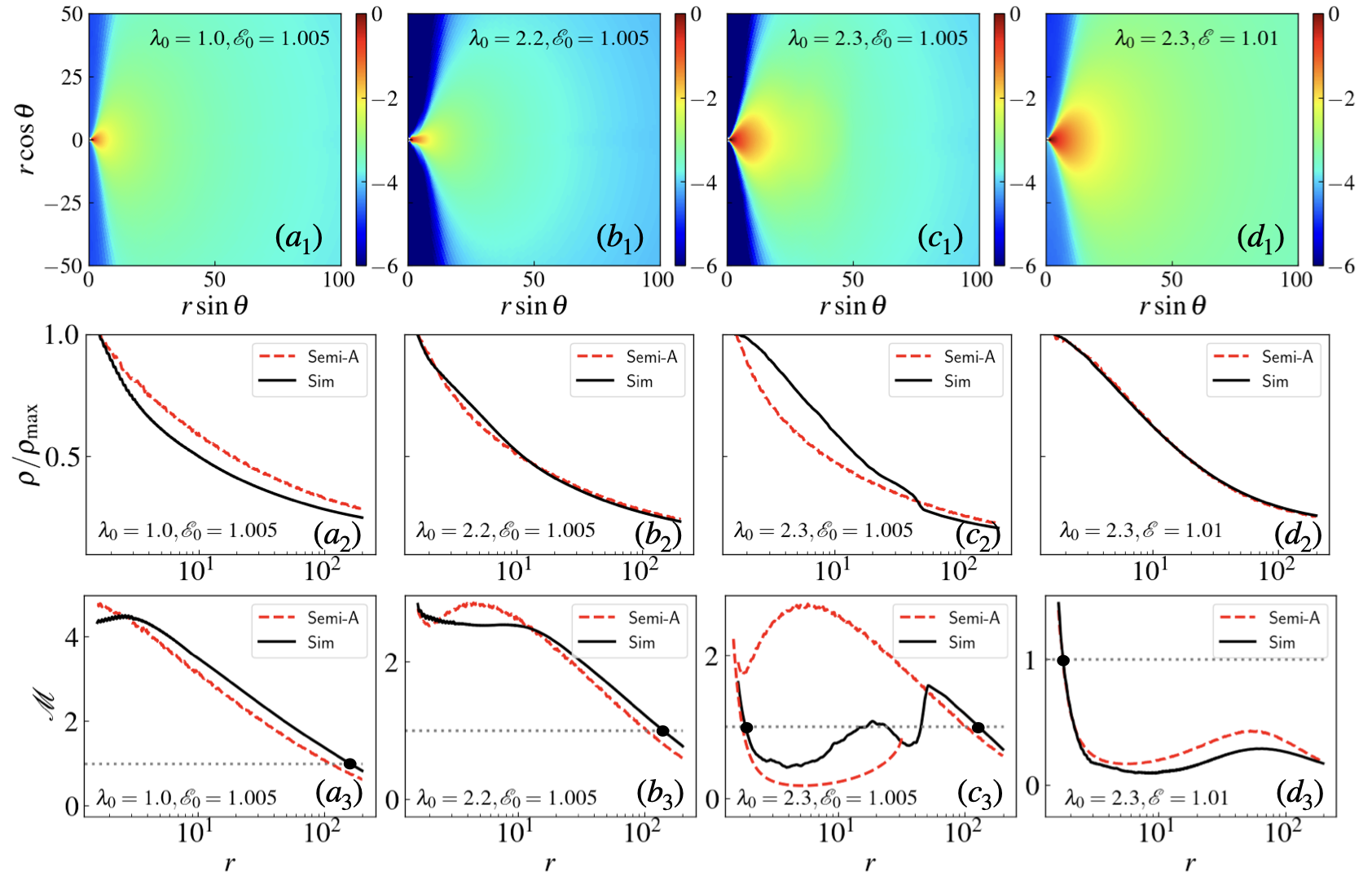}
    \caption{{\it Upper panels}: Plot of normalized logarithmic density ($\log(\rho/\rho_{\rm max})$) distribution on the poloidal plane. {\it Middle panels:} Plot of normalized density profile ($\rho/\rho_{\rm max}$) on the equatorial plane. {\it Bottom panels:} Plot of Mach number (${\cal M}$) profile on the equatorial plane. The values of specific angular momentum ($\lambda_0$) and energy (${\cal E}_0$) are written on every panel. Black solid and red dashed lines correspond to simulation and semi-analytical results, respectively. The solid circles in the Mach number profiles correspond to the sonic points. See the text for more details.}
    \label{fig:01}
\end{figure*}

In low-angular-momentum accretion flow, there are different types of accretion solutions. In order to show them, the upper, middle, and lower panels of Fig.~\ref{fig:01} show the normalized logarithmic density ($\log(\rho/\rho_{\rm max})$) distribution on the poloidal plane, the normalized density profile ($\rho/\rho_{\rm max}$), and the Mach number (${\cal M}=v/a_s$) profile on the equatorial plane. Here, $v$ is the radial velocity in the co-rotating frame, and $a_s$ is the local relativistic sound speed. The values of specific angular momentum ($\lambda_0$) and energy (${\cal E}_0$) are written on each panel. Additionally, the red dashed lines in the middle and lower panels correspond to the semi-analytical (Semi-A) solutions following \cite{Dihingia-etal2019a}. On the other hand, the solid black lines are obtained from the simulations (Sim). The horizontal lines on the lower panels indicate the Mach number ${\cal M}=1$, displaying the sonic point/points. 
In the simulations, we expect the values of $\lambda=-u_\phi/u_t$ and ${\cal E}=-hu_t$ to be conserved with their initial values, i.e., $\lambda_0$ and ${\cal E}_0$. We calculate the mean values for different panels in the region within $r\le500\,r_g$ and $\pm 10^\circ$ from the equatorial plane at the end of the simulation ($t=50\,000\,t_g$); they are given by $(a_1)~\lambda=1.0004, {\cal E}=1.0047$, $(b_1)~\lambda=2.2010, {\cal E}=1.0047$, $(c_1)~\lambda=2.3012,{\cal E}=1.0047$, and $(d_1)~\lambda=2.2969, {\cal E}=1.0114$, respectively. Thus, these values are very close to the initially given values (marked on each panel).

As discussed in semi-analytical studies, depending on the energy and angular momentum, we can have different types of accretion solutions. With our simulations, we also observe all different kinds of solutions: (i) passing through the inner sonic point (fourth column), (ii) passing through the outer sonic points (first and second columns), and (iii) passing through the two sonic points (third column). Follow the dotted horizontal line in the third row (${\cal M}=1$), to identify sonic points where flow transitions from subsonic to supersonic speed. 
The solid circles in the Mach number profiles in the third row correspond to the sonic points of the simulated results.
The transition towards the left is identified as the inner sonic point, and the transition towards the right is known as the outer sonic point. In the third panel, we see that there is one more transition from supersonic to subsonic in between (around $r_s\sim40-50\,r_g$), and it is much sharper than that of the other two transitions. These are identified as shock transitions. Since the shock location is changing in time, therefore, in the time-averaged plots we do not see a very sharp jump (see Appendix~\ref{App-B} for instantaneous sharp shock jumps). Instead, a diffused shock jump is shown. Our simulations confirm the formation of such shock transitions consistent with predictions from earlier semi-analytic studies of low-angular-momentum flows~\citep[e.g.,][etc.]{Chakrabarti1989,Das2007,Aktar-etal2015,Dihingia-etal2018,Dihingia-etal2018a,Dihingia-etal2019b}. Additionally, we find that our solutions from the simulations on the equatorial plane closely follow the semi-analytical solutions, as shown in the second and third rows by the red dashed lines. 

The simulations reproduce not only the equatorial plane behavior as the semi-analytical works do, but also provide all the important quantities in the off-equatorial plane. In order to calculate the observational signatures, the off-equatorial plane information is much more important than that of the equatorial behaviors (see Section~6 for more detail). From the density distribution, we can establish that the high-density region is more compact for solutions passing through the outer sonic points (first ($a_1$) and second ($b_1$) panels) since flow is supersonic far away from the event horizon of the black holes. For these cases, the sonic surface resides around $r\sim100\,r_g$. Whereas, for solutions having an inner sonic point, the density distribution is more extended. For solutions with two sonic points, the density distribution is more compact than that of a single inner sonic point but more extended than that of outer sonic point solutions. This is because subsonic flow is slower than that of its supersonic counterpart, and it can heat up the accretion flow, which could have distinct observational signatures. We will discuss it in the upcoming sections.

\begin{figure*}[ht]
    \centering
    \includegraphics[width=1.0\linewidth]{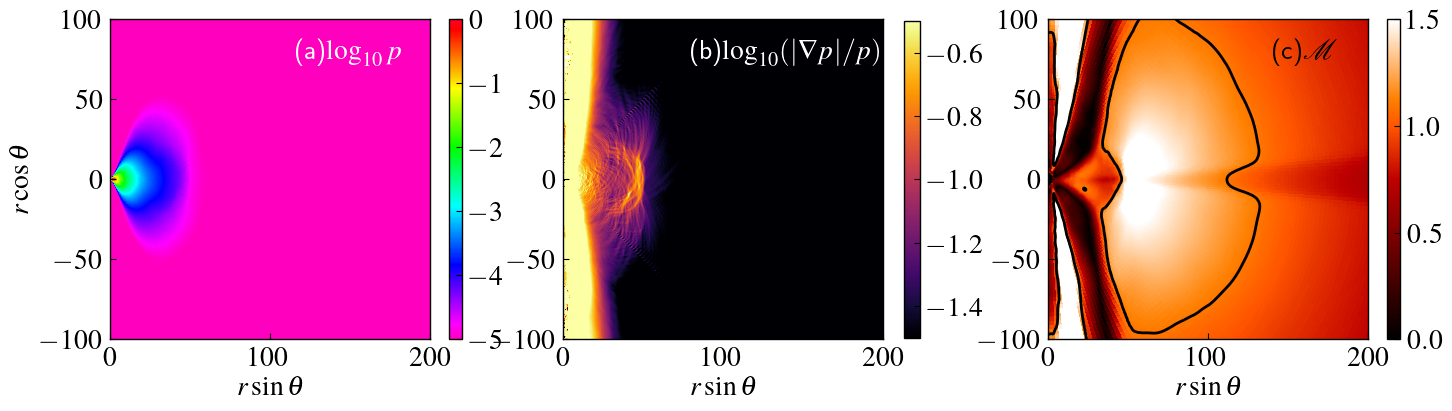}
    \caption{Distribution of (a) logarithmic pressure ($\log_{\rm 10}p$), (b) logarithmic normalized pressure gradient ($\log_{10}|\nabla p|/p$), and (c) Mach number (${\cal M}$) for $\lambda_0=2.3$ and ${\cal E}_0=1.005$. The solid contour in panel (c) corresponds to the boundary for ${\cal M}=1$.}
    \label{fig:02A}
\end{figure*}

Further to analyze the shock transition, in Fig.~\ref{fig:02A}, we show distribution of (a) pressure ($\log_{\rm 10}p$), (b) normalized pressure gradient ($\log_{\rm 10}|\nabla p|/p$), and (c) Mach number (${\cal M}$) for third column of Fig.~\ref{fig:01}, i.e., with $\lambda_0=2.3$ and ${\cal E}_0=1.005$. The solid contour in Fig.~\ref{fig:02A}(c) corresponds to sonic surface, i.e., ${\cal M}=1$. Fig.~\ref{fig:02A}(a) and \ref{fig:02A}(b) suggest that the post-shock region constitutes of a high-pressure region, and the pressure gradient has a sharp discontinuity at the supersonic to subsonic transition, which is a typical signature of shock jumps. Additionally, we see that the pressure gradient along the bipolar direction as well as in the post-shock region is significantly higher as compared to the pre-shock region. Due to this high pressure gradient, we expect to see significant outflow/jet from the post-shock region (see next section for more details). Finally, in Fig.~\ref{fig:02A}(c), we see the structure of the sonic surfaces. We clearly see the supersonic region in between the outer sonic surface (smooth transition) and the shock surface (with sharp discontinuity closer to the equatorial plane). Moreover, we observe a transition from subsonic to supersonic in the bipolar region, which suggests the presence of supersonic outflows/jets close to the black hole. The supersonic outflows/jet region can be seen with bright colors in the bipolar direction in Fig.~\ref{fig:02A}(c).

\begin{figure*}[ht]
    \centering
    \includegraphics[width=1.0\linewidth]{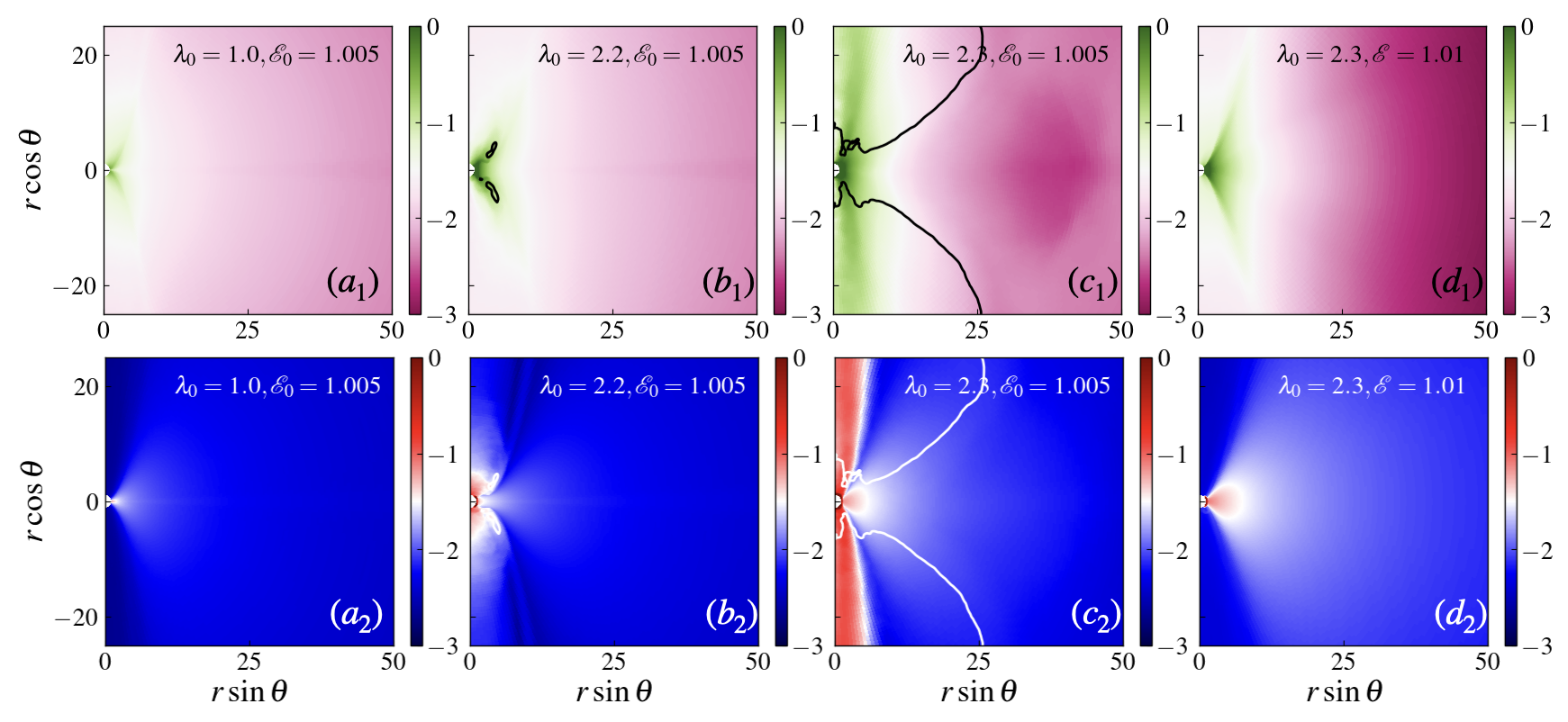}
    \caption{{\it (Upper panels)} distribution of Lorentz factor ($\log_{10}(\gamma-1)$) and {\it (lower panels)} distribution of temperature ($\Theta=p/\rho$) in the same fashion as the first row of Fig.~\ref{fig:01}. The solid (black/white) lines correspond to $u^r=0$. See the text for more details.}
    \label{fig:02}
\end{figure*}

\section{Possibilities of Outflows}
In this section, we study the impacts of types of accretion solutions on outflows (jets and winds). To study it, in Fig.~\ref{fig:02}, we show the distribution of the Lorentz factor ($\log_{10}(\gamma-1)$) and temperature ($\Theta=p/\rho$), respectively, in the same order as Fig.~\ref{fig:02}. The solid lines (white/black) in the panels correspond to the boundary of the surface $u^r=0$, separating inflow and outflow. 

With the increase of angular momentum, we see the formation of an outflow region close to the black hole in the third column ($\lambda_0=2.3, {\cal E}_0=1.005$). With increasing specific energy (fourth column), we do not see such outflows. Thus, the low-angular-momentum flow is largely inflow-dominated, unless it has a transition from two sonic points. Only in limited cases, we see the formation of outflows. It is evident from the first row that the outflow could have a Lorentz factor of the order of $\gamma\sim2$ close to the black hole. Such Lorentz factors are typically reported for black hole X-ray binary systems \citep[e.g.,][]{Fender-etal2004,Markoff2010, Belloni2010}. 

In this study, we do not invoke any magnetic fields; therefore, the launching jet/wind must be due to the thermal gradient force (see Fig.~\ref{fig:02A}(c) and supported by the centrifugal force. In the lower panels of Fig.~\ref{fig:02}, we observe that, depending on the types of solutions, the temperature profiles of the accretion flow are very different. For the solutions passing through the outer sonic points (first ($a_1$) and second ($b_1$) panels), we observe low-temperature accretion flow as compared to solutions passing through the inner sonic points (second ($c_1$) and third ($d_1$) panels). On the contrary, for solutions passing through both the sonic points, we see very hot bipolar regions, where the Lorentz factor is significantly higher and the density is lower. Thus, outflow shows two components: (i) low-density, hot, semi-relativistic jet and (ii) high-density, cold, non-relativistic winds. Note that, with the inclusion of strong magnetic fields, this scenario could be different. With magnetic fields, the semi-relativistic jets could accelerate to the relativistic regime \citep[e.g.,][]{Tchekhovskoy-etal2011,Porth-etal2017,Porth-etal2021}. However, with low-angular-momentum flow due to the shorter inflow time, it is challenging to incorporate magnetic fields consistently. We plan to do such studies in the future. 

\begin{figure*}[ht]
    \centering
    \includegraphics[width=0.8\linewidth]{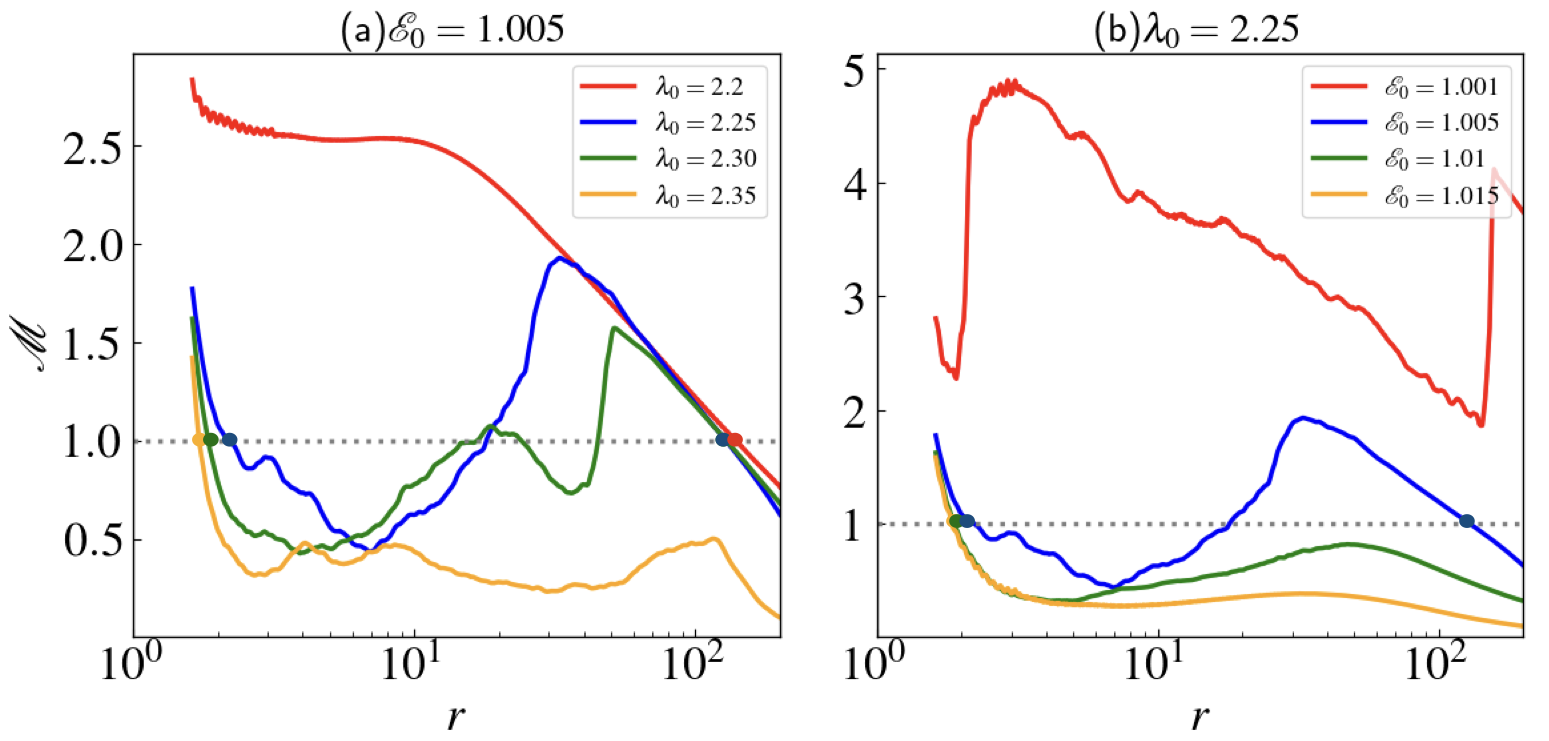}
    \caption{{\it Left}: radial distribution of time-averaged Mach number profiles with specific angular momentum ($\lambda_0$). {\it Right}: radial distribution of time-averaged Mach number profiles with specific energy (${\cal E}_0$) on the equatorial plane. The sonic points are marked with solid circles.}
    \label{fig:03}
\end{figure*}

\section{Dependence on Energy and Angular Momentum}
It is evident from Fig.~\ref{fig:01} that the solutions indeed depend on the choice of specific energy (${\cal E}_0$) and angular momentum ($\lambda_0$), which have been shown extensively in many previous semi-analytical studies \citep[e.g.,][]{Dihingia-etal2018a,Dihingia-etal2019a}. To investigate it, in Fig.~\ref{fig:03}, radial distributions of the time-averaged Mach number (${\cal M}$) on the equatorial plane are shown by varying $\lambda_0$ (left) and ${\cal E}_0$ (right), respectively. The sonic points are shown in the panels by solid circles with respective colors. We see that with the decrease of specific angular momentum, a solution with an inner point changes into solutions with two sonic points. Finally, with further decreasing the specific angular momentum, we find solutions passing through an outer sonic point. Similar to the variation of angular momentum, with the decrease of specific energy, we find similar variations in the accretion solutions. Thus, the number of sonic points in an accretion solution depends on the specific energy and angular momentum, which we confirm with our extensive GRHD simulations. Note that, in panel Fig.~\ref{fig:03}b (red curve), the accretion solution has reached a quasi-steady state only up to $r \sim 160\,r_g$. The outside region has not fully evolved yet for this simulation time(see Appendix ~\ref{App-B} for more detail). Lower energy solutions take more time to reach quasi-steady state than that of higher energy solutions. 

\section{Validation with 3D simulations}
\begin{figure*}
    \includegraphics[width=0.95\textwidth]{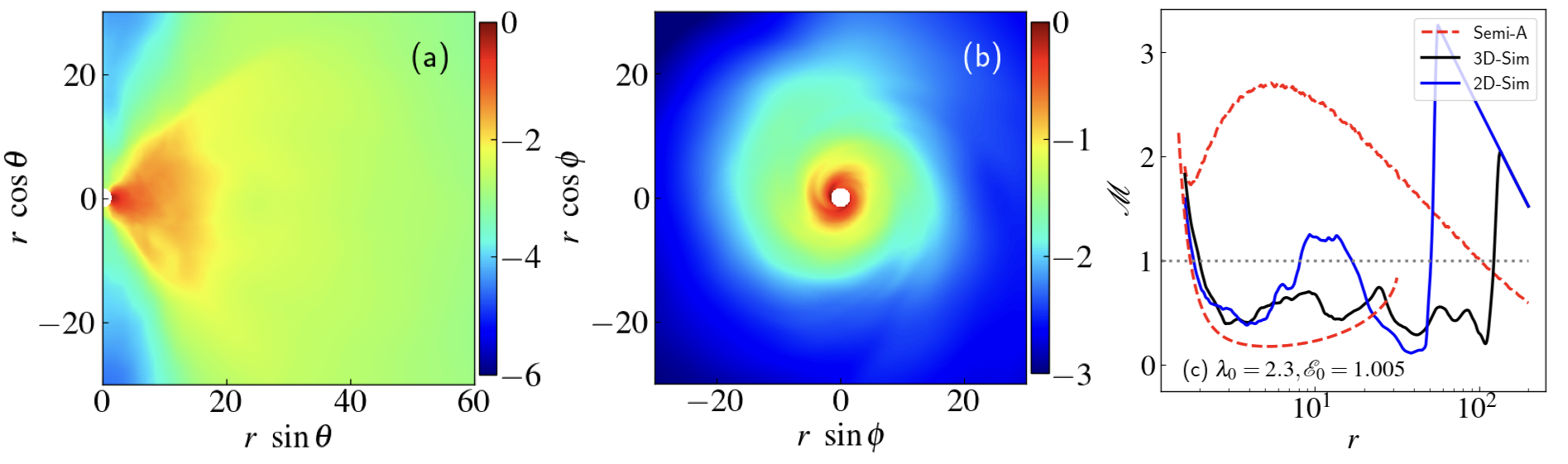}
    \caption{Time-averaged normalized logarithmic density ($\log_{10}(\rho/\rho_{\rm max})$) distribution of the poloidal (a) and equatorial plane (b) for 3D simulation with $\lambda_0=2.3$ and ${\cal E}_0=1.005$. (c) Mach number (${\cal M}$) for the same is shown on the equatorial plane along with the semi-analytical expectation by solid black and red-dashed lines, respectively. The solid blue line corresponds to the results obtained from the 2D simulation during the same simulation time range. The horizontal dotted line corresponds to ${\cal M}=1$.}
    \label{fig:05}
\end{figure*}
So far, our simulations have been extensively performed in 2D. Thus, we would like to validate our 2D simulation results with a 3D simulation. For that, we consider the case with $\lambda_0=2.3$ and ${\cal E}_0=1.005$ as a representative model and perform the simulations with a resolution of $256\times128\times64$. Since it is numerically expensive to run 3D simulations for a longer time, we performed it only up to $t=5000\,t_g$, and the time averaging is shown for $t=4500-5000\,t_g$ in Fig.~\ref{fig:05}. In panels Fig.~\ref{fig:05}a and Fig.~\ref{fig:05}b, we show the density distribution on the poloidal and equatorial plane, respectively. Panel (a) shows a very similar distribution to Fig.~\ref{fig:01}c$_1$, validating our 2D simulations. On the other hand, panel (b) shows that the distribution on the equatorial plane is very symmetric with respect to the black hole rotation axis. This makes our 2D simulations equally effective, robust, and reliable with 3D simulations. Finally, in Fig.~\ref{fig:05}c, we compare the Mach number profile with the semi-analytical and 2D simulations. We find that for this simulation time, the Mach number follows the solution passing through the inner sonic point. The inner sonic points obtained from both 2D and 3D simulations are located at the same radial distance.
We need more simulation time to reach the simulated Mach number in the solution passing through the outer sonic point. This is primarily due to the choice of initial disc height based on the analytical solutions. However, in future studies, if we choose a proper disc height based on current simulations, we will be able to reach analytical solutions in comparatively much less simulation time than the current one.

\begin{figure*}
    \centering
    \includegraphics[width=0.95\linewidth]{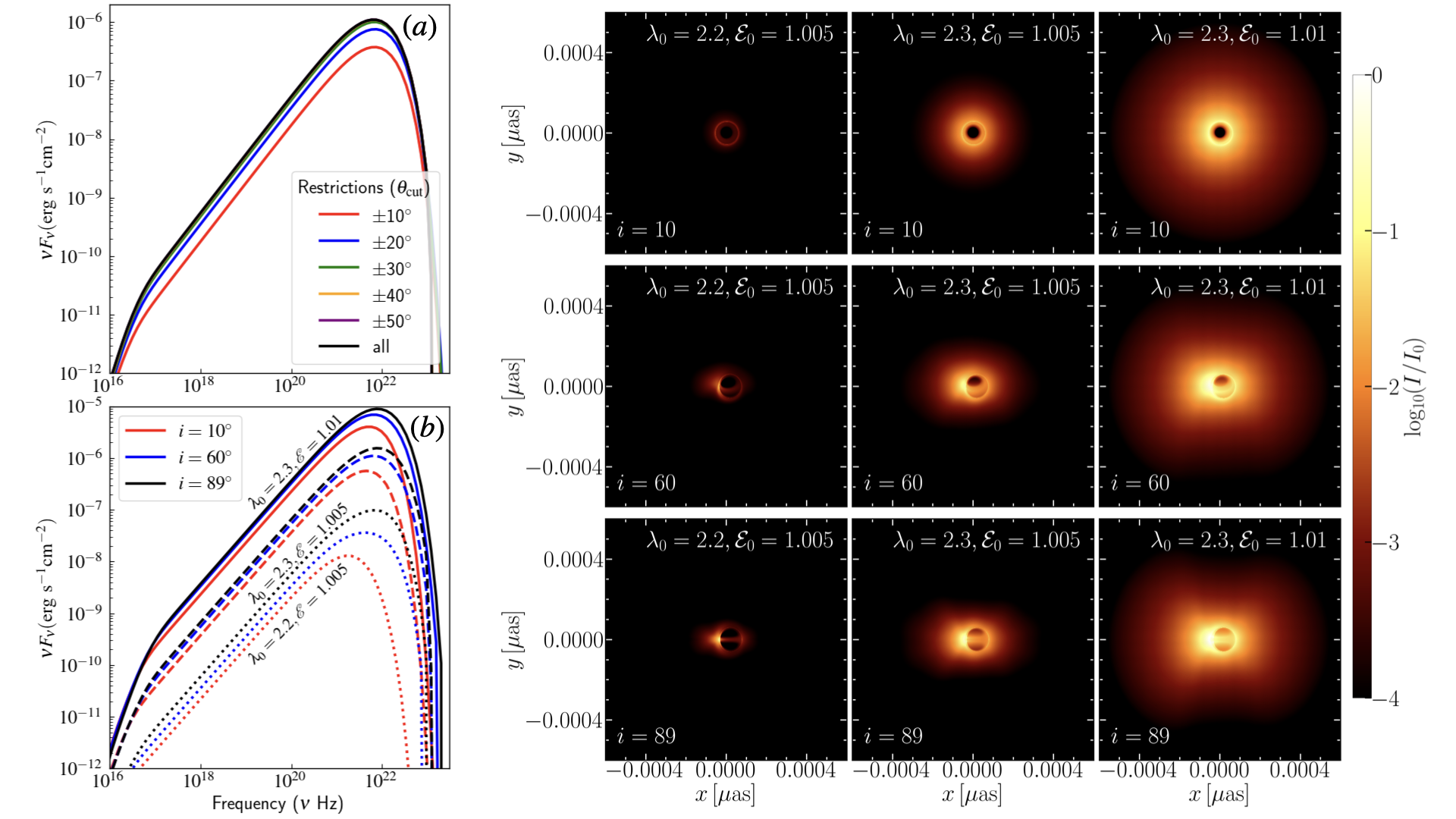}
    \caption{{\it (a) and (b) panels}: Spectral energy distribution of bremsstrahlung emission for different $\theta_{\rm cut}$ and different simulation models with different inclinations, respectively. {\it $3\times 3$ panels}: $1 \, \rm keV$ normalized logarithmic intensity ($\log_{10}(I/I_{\rm 0})$) maps for different specific energy (${\cal E}_0$) and angular momentum ($\lambda_0$) marked on the figures for three different inclination angles $i=30^\circ, 60^\circ,$ and $90^\circ$ in the upper, middle, and lower panels, respectively. The images are presented in micro-arcsecond ($\mu as$) scale.}
    \label{fig:04}
\end{figure*}

\section{Radiative signatures of different solutions}
In previous sections, we discussed the dynamics of accretion flows. Here, we study the impacts of types of solutions on the radiative properties. In order to do so, we consider bremsstrahlung emission from a source with mass $M_{\rm BH}=10M_\odot$ at a distance of $D=8$ pc. We utilized the general relativistic radiation transfer code {\tt RAPTOR} to perform these calculations~\citep{Bronzwaer-etal2018,Bronzwaer-etal2020}. We scale the mass of the accretion flow such that the total X-ray flux at $1\, \rm keV$ ($2.42\times10^{17}$Hz) is $10^{-9}$ erg cm$^{-2}$ s$^{-1}$. We search the mass unit value of density scaling for the case of ${\cal E}_0=1.01$ and $\lambda_0=2.3$, then we use the same density scaling for all the other models. Here, we use 2D hydrodynamic simulations only. Thus, we reconstruct a 3D flow structure assuming axisymmetry. This is a good approximation because the flow is mostly radial due to its low-angular-momentum nature. We use pixel resolutions of $500\times500$ to calculate the bremsstrahlung radiation. 

First, we investigate the impacts of the radiation coming from off-equatorial regions. For that, we set the restriction of the emission coming from $\theta_{\rm cut}=\pm 10^\circ, \pm20^\circ, \pm30^\circ, \pm40^\circ, \pm50^\circ$ and the full region (full). Fig.~\ref{fig:04}a shows the spectral energy distribution (SED) for the model with ${\cal E}_0=1.01$ and $\lambda_0=2.3$ with different restrictions. Here, we consider all emissions coming more than $\theta_{\rm cut}$ from the equatorial plane to be zero. The SEDs suggest that if we neglect the emission from the off-equatorial plane and consider the emission only from the equatorial region, the total emission starts to decrease significantly. It becomes prominent if the value of $|\theta_{\rm cut}|$ is less than $20^\circ$. 
It indicates that it is important to know the flow properties of the off-equatorial region. 
Otherwise, the emission properties calculated only from the 1D (equatorial) models would give inconsistent results, which is often applied in semi-analytic studies.

Hereafter, we consider the full region for the emission calculations.
In Fig.~\ref{fig:04}b, we show SEDs for three different cases with an outer sonic point ($\lambda_0=2.2,{\cal E}_0=1.005$), an inner sonic point ($\lambda_0=2.3,{\cal E}_0=1.01$), and both sonic points ($\lambda_0=2.3,{\cal E}_0=1.005$) in three different inclination angles ($i=10^\circ,60^\circ$, and $89^\circ$). The SEDs suggest that the emission decreases with decreasing inclination angle. We see maximum emission for the edge-on view (i.e., $i=89^\circ$). On the other hand, solutions with inner sonic points have the highest flux, whereas solutions with outer sonic points have the lowest flux. For the solutions with both sonic points has flux between other types of solutions. 

Finally, in the $3\times3$ panels, we show horizon-scale images at $1 \, \rm keV$ for these three inclination angles in the three cases considered. The values of $\lambda_0$, ${\cal E}_0$, and inclination angle $i$ are written on each panel. For better visualizations and comparisons, we scale the intensity in terms of maximum intensity for the case $\lambda_0=2.3,{\cal E}_0=1.01$, i.e., $I_0=4.28\times10^{-7}$erg cm$^{-2}$ s$^{-1}$. In the images, we see the photon rings quite similar to synchrotron emission images for supermassive black holes \citep[e.g.,][]{EHTC2019,EHTC2022,EHTM872025}. In these panels, we observe maximum intensity from the solution with the inner sonic point and the lowest intensity for the solution with the outer sonic point. The solution with both sonic points presents intermediate intensity. This result can be understood from the density and temperature distribution shown in Fig.~\ref{fig:01}b$_1$,c$_1$, d$_1$, and Fig.~\ref{fig:02}b$_2$,c$_2$, d$_2$, respectively. We observe a more extended density and a higher temperature for the solution with an inner sonic point than for that of other solutions. The high intensity arises from emissions in these extended, high-density, hot regions. 
Additionally, we see morphological differences among the three different solutions. The solutions with the outer sonic point make the most compact emission structure. On the other hand, the solutions with an inner sonic point create a largely extended emission. The solutions with both sonic points also show extended emission, but still smaller than that seen in the solution with the inner sonic point.
In summary, the broadband SED and horizon-scale emission structure significantly depend on the types of low-angular-momentum accretion solutions. 
Finally, we would like to mention that we will not be able to observe them with the Event Horizon Telescope or similar missions because of the very small aperture of the images and the limited resolution of these imaging missions. 
Nonetheless, they give a comprehensive understanding of the emission regions of the accretion flow. Additionally, in a realistic scenario (magnetized flows), synchrotron and inverse Compton (IC) scattering of synchrotron photons may be important depending on the energy ranges and magnetic field strengths. In the pseudo-Newtonian approach, \cite{Chatterjee-etal2017,Chatterjee-etal2018} show that the radiative properties (including IC) depend on the post-shock region. With GRMHD and GRRT, coupling dynamical plasma and radiation requires more explicit tests in the future. However, for current non-magnetized flow, inverse Compton scattering of bremsstrahlung photons will not alter the distinctive feature of images at $1 \, \rm keV$ for different types of solutions.

\section{Conclusions and discussions}
This work investigates three different types of low-angular-momentum accretion solutions with two-dimensional (2D) GRHD simulations, which we validated with the help of a three-dimensional (3D) simulation.
Additionally, we study radiative imprints of different types of solutions, considering bremsstrahlung emission from the accretion flow with GRRT calculations. We list our conclusions below:
\begin{itemize}
    \item[1.] We find that a transonic accretion solution can exhibit either one (inner or outer) or two sonic points (inner and outer), with a transition from supersonic to subsonic in between through shock phenomena. Shocks appear intermittent or oscillatory, and therefore, they can not be seen as sharp transitions in the time-averaged distributions/profiles. The time-averaged profiles of accretion flows on the equatorial plane closely follow the earlier semi-analytical solutions \citep[e.g.,][]{Dihingia-etal2018a,Dihingia-etal2019a}.

    \item[2.] With the change of specific energy (${\cal E})_0$) and angular momentum ($\lambda_0$), the solution can transit from one to another. The solutions with two sonic points have an additional signature of hot jets/outflows of Lorentz factor $\gamma\sim2$ without invoking any magnetic fields, which is not seen for single sonic point solutions. This reinforces the idea that thermal gradients may be sufficient to produce semi-relativistic outflows in certain parameter regimes.

    \item[3.] We find that the radiative properties of low-angular-momentum flow are also affected by the off-equatorial region. The dependency is prominent for $|\theta_{\rm cut}|<20^\circ$. Thus, a full 2D solution rather than a 1D solution is appropriate for investigating radiative signatures from such flows. 

    \item[4.] We can arrange the different types of accretion solutions in terms of ascending radiation flux as follows: (i) solutions with an outer sonic point, $<$ (ii) solutions with both sonic points $<$ (iii) solutions with an inner sonic point. 

    \item[5.] The horizon-scale images of $1 \, \rm keV$ suggest that it is more compact for solutions with an outer sonic point, whereas the images are largely extended for solutions with an inner sonic point. For the solutions with both the sonic points, the extension of the images is between them. We confirmed it at different inclination angles.
\end{itemize}

In summary, our simulations show two-dimensional structures of different types of low-angular-momentum accretion solutions, which we also validated with a short 3D simulation. We plan to do long 3D simulations in the future. Of the three solution types, only those with two sonic points exhibit hot jet/outflow signatures. Note that the jet/outflow is launched here without magnetic fields, which is due to thermal pressure. We observe a Lorentz factor of the order of $\gamma\sim2$. Such mildly-relativistic jets are often observed in BH-XRBs \citep[e.g.,][]{Fender-etal2004,Markoff2010, Belloni2010}. With the introduction of magnetic fields, we expect more powerful relativistic jets will be formed, which we plan to study in the future. 

We find that the radiative properties (SED, horizon-scale image) depend significantly on the types of accretion solutions. Since the equatorial plane solution closely follows semi-analytical models, it is possible to construct a full solution without rigorous numerical simulations, considering current simulations as a reference. With such construction, a full library of radio images ($230$ GHz) can be generated even for supermassive black hole sources (e.g., Sgr~A$^*$) considering weak magnetic field limits. 
From Event horizon telescope (EHT) studies, it has been established based on MAD (Magnetically Arrested Disk, i.e., strongly magnetized) and SANE (Standard and Normal Evolution, i.e., weakly magnetized) models, that MAD models are favored for M\,87 and Sgr A$^*$\citep{EHTC2019, EventHorizonTelescope:2019ggy, EventHorizonTelescope:2019ths, EventHorizonTelescope:2021srq, EventHorizonTelescope:2021bee, EHTC2022, EventHorizonTelescope:2022exc, EventHorizonTelescope:2022urf, EventHorizonTelescope:2024hpu, EventHorizonTelescope:2024rju, EHTM872025}; however, non-MAD models still need to be tested systematically.
Such a systematic study could give us opportunities to test low-angular-momentum accretion flows around Sgr~A$^*$, which is often suggested by earlier studies \citep{Ressler-etal2018,Ressler-etal2023,Dihingia-Mizuno2024}. Thus, chances are that the scenario for Sgr~A$^*$ may be altered. We plan to do such studies in the future.

\begin{acknowledgments}
This research is supported by the National Key R\&D Program of China (Grant No.\,2023YFE0101200), the National Natural Science Foundation of China (Grant No.\,12273022), the Research Fund for Excellent International PhD Students (grant No. W2442004) and the Shanghai Municipality orientation program of Basic Research for International Scientists (Grant No.\,22JC1410600). I.K.D. acknowledges the TDLI postdoctoral fellowship for financial support. The simulations were performed on the TDLI-Astro cluster in Tsung-Dao Lee Institute, Pi2.0, and Siyuan Mark-I clusters in the High-Performance Computing Center at Shanghai Jiao Tong University. This work has made use of NASA's Astrophysics Data System (ADS). We appreciate the thoughtful comments and suggestions provided by the anonymous reviewers that have improved the manuscript.
\end{acknowledgments}

\section*{Data Availability}
The simulation data and analysis scripts used in this work are available upon reasonable request. 

\appendix

\section{Semi-analytical solutions and initial conditions}\label{App-A}
By simple steps of calculations, with the help of the GRHD equations, the radial derivatives of the radial velocity on the equatorial plane in the corotating frame can be expressed as $dv/dr={\cal N}/{\cal D}$. For explicit expressions of the GRHD equations on the equatorial plane and ${\cal N}$ and ${\cal D}$, please follow \cite{Dihingia-etal2018a,Dihingia-etal2019a}. We solve the equations from the sonic point, where ${\cal N}={\cal D}=0$ simultaneously. Depending on the input parameters, viz., specific energy ${\cal E}_0=-hu_t$, and specific angular momentum $\lambda_0=-u_\phi/u_t$, we can have either one or three sonic points. At the sonic points, we use L'H\^{o}pital's rule to get the two values of the velocity gradient ($dv/dr|_c$). If the sign of both values of $dv/dr|_c$ is real and opposite, such points are known as saddle type or `X-type'. If they are of the same sign, sonic points are known as nodal type. Finally, if the values are imaginary, then the sonic points are known as `O-type'. Here, we only consider `X-type' sonic points with a `-ve' sign for $dv/dr|_c$, which corresponds to the accretion solution. We solve $dv/dr$ and $d\Theta/dr$ (temperature gradient: $\Theta=p/\rho$) starting from the sonic point towards both sides and joining them, we get the full accretion solution connecting the event horizon and infinity. After that, we use the solution $v(r)$ to get all the initial conditions as follows:
\begin{align}
\begin{aligned}
    u^r(r,\theta) &= -g_{rr}^{1/2}\frac{v(r)}{\sqrt{1-v^2(r)}}f(\theta), ~~~~
    u^\theta(r,\theta)  = 0,\\
    u^\phi(r,\theta) & = g^{\phi\phi}u_\phi + g^{t\phi}u_t,~~~~
    u^t(r,\theta)  = g^{tt}u_t + g^{t\phi}u_\phi,\\
    \rho(r,\theta) & = \bigg[\frac{\Gamma - 1}{\kappa\Gamma}(h_0 - 1)\bigg]^\frac{1}{\Gamma - 1},~~~~
    p(r,\theta)  = \kappa \rho^\Gamma,\\
\end{aligned}
\end{align}
where
\begin{align}
    h_0 (r,\theta) =  \sqrt{\frac{( 2\lambda_0g^{t\phi} - g^{tt}-\lambda_0^2g^{\phi\phi}){\cal E}_0^2}{1 + g_{rr}(u^r)^2}},~~
    u_t(r,\theta)  = -{\cal E}_0/h_0,~~
     u_\phi(r,\theta)  = -\lambda_0u_t.
\end{align}
Here $f(\theta)$ is an assumed function, which models the distribution along the vertical direction. For simplicity, we consider that $u^r$ increases in the vertical direction depending on a scale height, which is given by
\begin{align}
    f(\theta) =
\begin{cases}
1 + \sin^2(\frac{\pi\cos\theta}{\cos\theta_H}), & \text{if } |\cos\theta| \le |\cos\theta_H|/2 \\
0, & \text{otherwise.}
\end{cases}
\end{align}
Here $H=r\cos\theta_H$ gives the initial disk height obtained from the semi-analytical calculations (following eq. 18 of \cite{Dihingia-etal2018a}).

\section{Time variation of solutions}\label{App-B}
\begin{figure*}
    \centering
    \includegraphics[width=0.99\linewidth]{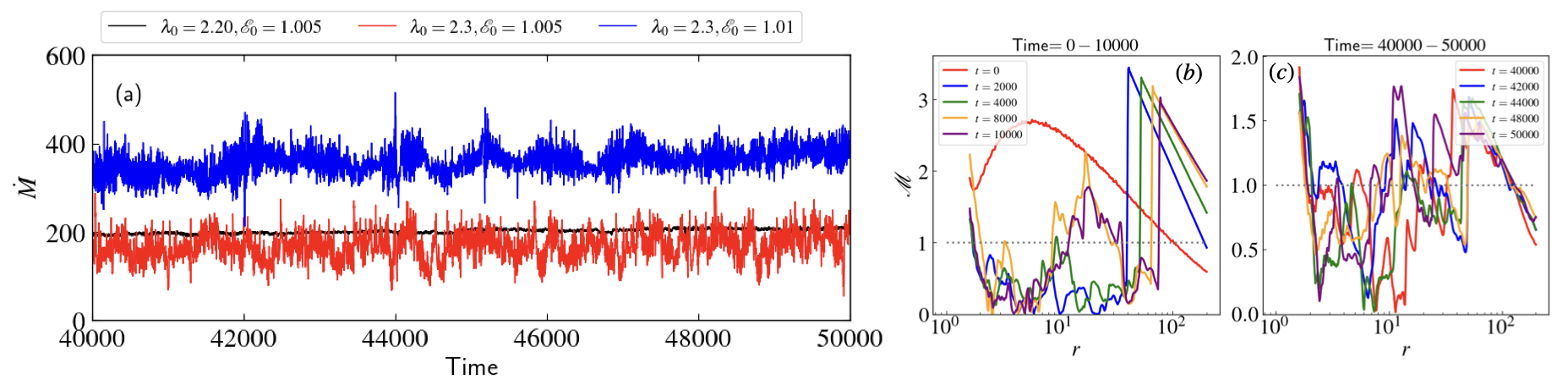}
    \caption{{\it Panel (a).} Profiles of mass accretion rate at the event horizon for three different models with different specific energy (${\cal E}_0$) and angular momentum ($\lambda_0$). {\it Panels (b) and (c).} Mach number profiles (${\cal M}$) calculated for $\lambda_0=2.3, {\cal E}_0=1.005$ for different simulation time ranges $0-10000\,t_g$ and $40000-50000\,t_g$, respectively. The dotted horizontal line corresponds to the Mach number ${\cal M}=1$.}
    \label{fig:a1}
\end{figure*}

In this section, we study the time variation of low-angular-momentum flow. Accordingly, in Fig.~\ref{fig:a1}a, we show the accretion rate ($\dot{M}$) at the event horizon in simulation time $t=40\,000-50\,000\,t_g$ for three different types of solutions passing through (i) the outer sonic point ($\lambda_0=2.2, {\cal E}_0=1.005$, black), (ii) the inner sonic point ($\lambda_0=2.3, {\cal E}_0=1.005$, red), and (iii) both the sonic points ($\lambda_0=2.3, {\cal E}_0=1.01$, blue), respectively. We observe that during these simulation times for all the cases, the accretion rate reaches a quasi-steady state. Additionally, in panels Fig.~\ref{fig:a1}b and Fig.~\ref{fig:a1}c, we show the instantaneous Mach number profiles on the equatorial plane for different times in between $t=0-10\,000\,t_g$ and $t=40\,000-50\,000\,t_g$, respectively. 
The corresponding simulation times are written in each panel. During $t=0-10\,000\,t_g$, we see some sharp transitions in the Mach number, and the location of the transition moves outward from the black hole with time. This is because of the inconsistent initial conditions of the off-equatorial plane. More precisely, we fill the initial simulation domain in the region $|\cos\theta|\le|\cos\theta_H|/2$. With time, matter starts to occupy the off-equatorial region, creating a quasi-steady distribution over time. This sharp jump could often be mistaken for a shock front, and interestingly, it can be made steady/oscillatory by giving initial conditions suitably. 
In order to find an appropriate physical shock location, one must check for a shock strength ${\cal S}\simeq{\cal M_-}/{\cal M_+} = {\cal M}_-({((\Gamma - 1) {\cal M}_-^2 + 2)}/{(2\Gamma {\cal M}_-^2 - (\Gamma - 1))})^{-1/2}$ for Rankine-Hugoniot shock jumps, where `+' and `-' correspond to post-shock and pre-shock quantities, respectively. The values of ${\cal S}$ are usually much higher for these artifacts (unphysical shock). For example, at $t=10\,000\,t_g$, we see a value of ${\cal S}\sim9$ for the artifact, but it should be ${\cal S}\sim6.5$ according to the Rankine-Hugoniot shock jump conditions. Similarly, the compression ratio should also be within ${\cal R}\simeq\rho_+/\rho_-={(\Gamma + 1) {\cal M}_-^2}/{(\Gamma - 1) {\cal M}_-^2 + 2}\le\Gamma+1/\Gamma-1\le7$ for Rankine-Hugoniot shock jump conditions with $\Gamma=4/3$.
Note that the given function $f(\theta)$ is chosen in such a way that we could reach a quasi-steady distribution faster. Ideally, a constant distribution also gives similar results, but the time to reach quasi-steady state becomes longer. The final distribution is expected to be independent of the choice of $f(\theta)$. 

Finally, Fig.~\ref{fig:a1}c shows that the Mach number profiles during time $t=40\,000-50\,000\,t_g$ are more or less similar, suggesting a quasi-steady behavior of the flow. In the profiles, we can identify the first cut of ${\cal M}=1$ lines towards the left and right as inner sonic points and outer sonic points, respectively. We observe that these two points remain mostly in the same position throughout the simulation. On the other hand, in between these two points, we observe more crossing on ${\cal M}=1$ lines; all these crossings disappear in the time-averaged profile except the crossing around $r=40-50\,r_g$. The transition around $r=40-50\,r_g$ is sharp and connects the solutions passing through the inner sonic point and outer sonic point. Accordingly, we identify it as a shock transition as proposed by many semi-analytical studies \citep[e.g.,][]{Dihingia-etal2018a,Dihingia-etal2019a}. Note that the shock strength calculated for such shocks follows the Rankine-Hugoniot shock jump conditions, with a value of ${\cal S}\sim2.8$. Nonetheless, more focused studies and more simulations are needed to characterize this transition, which we plan to do in the future. 
\bibliography{sample7}{}
\bibliographystyle{aasjournalv7}



\end{document}